\documentclass[lettersize,journal]{IEEEtran}
\usepackage{amsmath,amsfonts}
\usepackage{algorithmic}
\usepackage{algorithm}
\usepackage{array}
\usepackage[caption=false,font=normalsize,labelfont=sf,textfont=sf]{subfig}
\usepackage{textcomp}
\usepackage{stfloats}
\usepackage{url}
\usepackage{verbatim}
\usepackage{graphicx}
\usepackage{cite}
\usepackage{amssymb}
\usepackage{graphicx}
\usepackage{xcolor}
\usepackage{booktabs}
\usepackage{adjustbox}
\usepackage{multirow}
\usepackage{hyperref}
\usepackage{cleveref}

\begin{document}

\title{SALMONN-2: Advancing General-Purpose Hearing Abilities with Self-Supervised Representations}

\author{
Xiaoyu Yang$^\ast$,
Xuenan Xu$^\ast$,
Wenyi Yu,
Siyin Wang,
Changli Tang,
Terumi Chiba,
Siyuan Hou,
Ziyang Zhang,
Wen Wu,
Baoxiang Li,
Guangzhi Sun,
Chao Zhang$^\dagger$,
Philip Woodland
\thanks{$^\ast$Xiaoyu Yang and Xuenan Xu contributed equally to this work.}
\thanks{$^\dagger$Chao Zhang is the corresponding author.}
\thanks{Xiaoyu Yang, Guangzhi Sun, and Philip Woodland are with the University of Cambridge.}
\thanks{Xuenan Xu, Wen Wu, Baoxiang Li, and Chao Zhang are with the Shanghai Artificial Intelligence Laboratory.}
\thanks{Wenyi Yu, Siyin Wang, Changli Tang, Terumi Chiba, Siyuan Hou, Ziyang Zhang, and Chao Zhang are with Tsinghua University.}
\thanks{This work has been submitted to the IEEE for possible publication. Copyright
may be transferred without notice, after which this version may no longer be
accessible.}
}

\markboth{Submitted to IEEE Transactions on Audio, Speech, and Language Processing}%
{Yang \MakeLowercase{\textit{et al.}}: SALMONN-2}


\maketitle

\begin{abstract}

Recent audio large language models (ALLMs) are typically built upon audio encoders trained with large amounts of supervised data.
Since self-supervised learning (SSL) audio encoder models are known to learn general-purpose and transferable representations, we investigate whether general-purpose SSL audio representations can serve as an effective foundation for ALLMs.
We present SALMONN-2, an ALLM built upon a unified SSL encoder.
To better exploit the hierarchical representations learned by SSL encoders, we propose a multi-layer feature fusion (MLF) adapter that aggregates information from all encoder layers before projecting them into the language model.
Beyond conventional audio understanding tasks, we further explore multimodal in-context learning (MICL) in ALLMs and study how this capability can be acquired through contextual biasing training.
Experimental results show that a general-purpose SSL encoder achieves performance comparable to, or better than, specialised supervised audio encoders while providing a more balanced capability across speech, audio, music and paralinguistic tasks.
SALMONN-2 further achieves state-of-the-art performance among comparable-scale open-weight models on ALLM understanding benchmarks, obtaining the best results on MMAU-Pro, MMAR and MMSU.
We also show that MICL does not emerge naturally in ALLMs, but can be effectively acquired through targeted contextual biasing training.
\end{abstract}

\section{Introduction}

Recent advances in large language models (LLMs) have accelerated the development of multimodal systems capable of understanding and reasoning over diverse input modalities.
Among them, audio large language models (ALLMs) have emerged as a promising paradigm for general-purpose audio intelligence, enabling audio understanding~\cite{salmonn,qwen2audio}, reasoning~\cite{mimo-audio,step-audio-r1.5}, and generation~\cite{qwen2.5-omni,kimi-audio} capabilities.
As an early effort, SALMONN~\cite{salmonn} demonstrates that LLMs can be extended to understand speech, audio events, and music through dual-encoder audio perception and audio-language instruction tuning.

Although SALMONN establishes a general framework for connecting audio perception modules with LLMs, the design of several key components in ALLMs, including the choice of audio encoder, the connector between audio encoder and LLM, and the utilisation of contextual information, remains underexplored.
Despite encouraging advances in recent ALLMs, most of them inherit similar architectural designs, with efforts focused on scaling model size and instruction data, whereas comparatively less attention has been paid to these fundamental design choices.

Acoustic signals contain not only linguistic content but also non-linguistic information, such as speaker characteristics, emotions, music, and environmental sounds.
Consequently, the effectiveness of audio representations is critical to the capabilities of ALLMs.
Most existing ALLMs are built upon supervised audio encoders trained on large-scale annotated data.
However, supervised learning is inherently constrained by the quality, diversity, and coverage of annotations.
As a result, the learned representations may fail to capture acoustic information beyond the scope of pre-training tasks.
In contrast to supervised learning, self-supervised learning (SSL) leverages large-scale unlabelled audio data to capture rich acoustic information, resulting in more generalisable and transferable audio representations~\cite{spear}.
To cover diverse acoustic information, some ALLMs combine multiple audio encoders~\cite{salmonn,kimi-audio}.
These designs raise a fundamental question: can a single general-purpose SSL-based audio encoder serve as a unified and effective audio perception module for ALLMs?

Beyond the choice of encoder, effectively integrating audio representations into an LLM is another important challenge.
Existing ALLMs typically employ simple adapters that project the final-layer representations of the audio encoder into the LLM embedding space.
This design overlooks the hierarchical representations learned by audio encoders: early layers capture low-level acoustic information while deeper layers encode high-level semantic concepts.
How to leverage such rich hierarchical representations through the adapter remains underexplored.

\begin{figure}[ht]
    \centering
    \includegraphics[width=\linewidth]{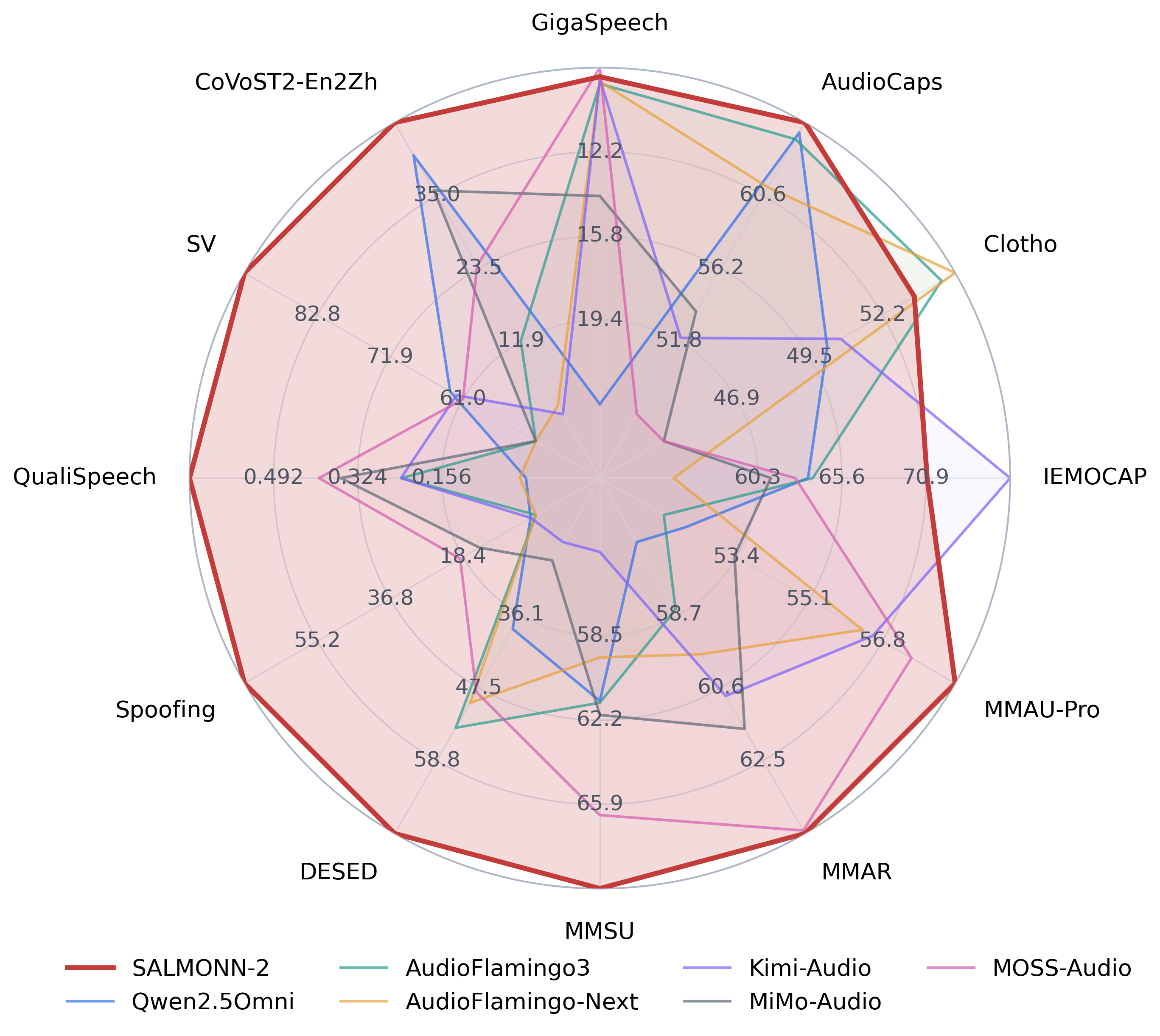}
    \caption{Performance of SALMONN-2 and existing open-weight ALLMs. SALMONN-2 achieves competitive performance on audio understanding tasks (ASR, AAC, ER, etc.) and benchmarks (MMSU, MMAR and MMAU-Pro), while exhibiting new capabilities on audio analysis tasks that were previously overlooked, such as speech quality assessment, SED, and spoofing detection.}
    \label{fig:radar_chart}
    \vspace{-0.7em}
\end{figure}

In addition to the representation and adapter design, another underexplored aspect of ALLMs is the utilisation of contextual information.
The success of textual LLMs is largely attributed to their ability to perform in-context learning (ICL), enabling adaptation to new domains or tasks through context, without extra post-training.
For audio understanding, contextual information, especially multimodal context, also plays an important role.
For example, in contextualised automatic speech recognition (ASR), domain-specific phrases or hotwords and their pronunciations provide informative cues to improve recognition accuracy.
In contrast, existing ALLMs rarely investigate ICL for audio understanding, especially multimodal in-context learning (MICL), where audio-text demonstrations are provided as contextual examples.
As a result, the powerful ICL capabilities of LLMs are not fully exploited in current ALLMs.

In this work, we substantially extend SALMONN to propose SALMONN-2 and provide a systematic study of the aforementioned designs: audio representations, hierarchical acoustic feature adaptation, and multimodal in-context learning for ALLMs.
SALMONN-2 is built upon an SSL audio encoder, a multi-layer feature fusion (MLF) adapter, and MICL training.
SPEAR~\cite{spear} is adopted as the audio encoder, which provides strong and generalisable audio representations across diverse domains and tasks.
To better exploit the hierarchical representations learned by SPEAR, we introduce a simple yet effective multi-layer feature fusion adapter that integrates information from all encoder layers before projecting audio features into the LLM embedding space.
Experimental results demonstrate that the combination of a single SPEAR encoder and the MLF adapter forms an effective audio perception module, outperforming multi-encoder designs while avoiding additional architectural complexity.
Furthermore, SALMONN-2 is trained with explicit MICL instruction-tuning data, enabling it to effectively leverage contextual information.
Despite training on short contextual samples only, SALMONN-2 exhibits an emerging capability to generalise to longer multimodal contexts.

Our main contributions are summarised as follows:
\begin{itemize}
\item We substantially extend SALMONN to a more systematic study of audio perception, audio-language adaptation, and contextual learning for ALLMs. This leads to SALMONN-2, a stronger and more efficient general-purpose ALLM.
\item We investigate whether a single general-purpose SSL model can serve as a unified audio perception module for ALLMs. Together with the proposed MLF adapter, the SPEAR encoder consistently improves downstream performance and is comparable to or better than multi-encoder designs.
\item We investigate MICL in ALLMs and demonstrate that explicit MICL training enables effective utilisation of multimodal contextual information, outperforming text-only ICL.
\item SALMONN-2 achieves state-of-the-art performance among similar-scale open-weight models across diverse audio understanding benchmarks, including MMAU-Pro, MMAR and MMSU with a highly efficient training setup with fewer than 20K hours of instruction-tuning data (see \Cref{fig:radar_chart}). Furthermore, we demonstrate that scaling the LLM backbone from 8B to 30B-A3B consistently improves performance on challenging audio understanding benchmarks, highlighting the scalability of our framework. We open-source SALMONN-2 to facilitate future ALLM research\footnote{Code and model are available at \url{https://github.com/bytedance/SALMONN/tree/salmonn2}.}.
\end{itemize}







\section{Related Work}

\subsection{Audio Large Language Models}

ALLMs aim to extend the perception and reasoning abilities of text-based LLMs. By connecting an audio encoder to a text LLM, ALLMs accept audio as input and perform audio understanding or generation tasks leveraging the LLM decoder~\cite{salmonn,qwenaudio,qwen2audio,midashenglm,qwen2.5-omni}.
Early systems such as SALMONN~\cite{salmonn} adopt a dual-encoder architecture, combining a speech-oriented encoder and a general audio encoder to cover both speech and non-speech audio understanding. More recent ALLMs move towards unified audio encoders, often trained on large amounts of supervised audio-text data~\cite{mimo-audio, audioflamingo3}.

Another important design choice is the connector between the audio encoder and the LLM.
Existing ALLMs commonly use linear projection modules~\cite{qwenaudio}, Q-Former-style architectures~\cite{salmonn}, or other adapter networks to map audio features into the LLM embedding space. However, most systems directly use the final-layer encoder representation or treat the encoder output as a single feature stream, leaving the use of layer-wise encoder representations insufficiently studied.
MOSS-Audio~\cite{moss-audio}, a concurrent work, proposes to inject audio features from multiple encoder layers into multiple LLM decoder layers.
In contrast, SALMONN-2 proposes a simpler yet effective fusion mechanism to exploit hierarchical encoder representations without intruding into the text LLM.



\vspace{-0.8em}
\subsection{Unified Self-Supervised Audio Models}

SSL has been widely used to learn general-purpose representations from unlabelled data, achieving strong results across speech~\cite{hubert, chen2022wavlm}, general audio~\cite{audiomae, beats}, and music~\cite{mert, muq} domains. However, many prior SSL models are designed for a specific domain, which limits their applicability to handle the heterogeneous acoustic information required by ALLMs.
Recent work has begun to extend SSL audio modelling towards unified representations across multiple domains. USAD~\cite{usad} utilises feature-level knowledge distillation to train a single encoder to imitate two domain-specific SSL teachers, covering speech and general audio respectively.
In contrast, SPEAR~\cite{spear} develops a unified SSL framework for speech and general audio by combining teacher knowledge distillation with a masked-token prediction objective, achieving strong results on multiple speech and audio benchmarks.

These unified SSL models demonstrate that a single encoder can capture rich acoustic information across domains.
However, their potential as audio encoders for ALLMs remains underexplored, especially in terms of how their representations can be integrated into an LLM.
SALMONN-2 adopts SPEAR as a unified audio encoder and investigates how to effectively exploit its hierarchical representations for general-purpose audio language modelling.


\vspace{-0.6em}
\subsection{Multimodal In-context Learning in ALLMs}

Recent ALLMs have started to explore multimodal in-context learning in the audio domain.
Audio Flamingo~\cite{audioflamingo} explicitly trains models for few-shot learning with paired audio-text data and retrieval-augmented inference.
MiMo-Audio~\cite{mimo-audio} further studies few-shot capabilities in speech generation scenarios such as voice conversion.
In contrast, MICL for contextual speech recognition remains underexplored.
Contextual ASR is a particularly suitable setting for studying MICL, since external context can directly affect the recognition of rare words and named entities.
SALMONN-2 investigates MICL in contextual ASR and shows that such capability requires targeted training rather than emerging automatically from standard audio-text instruction tuning. 


\begin{figure*}[ht]
    \centering
    \includegraphics[width=\linewidth]{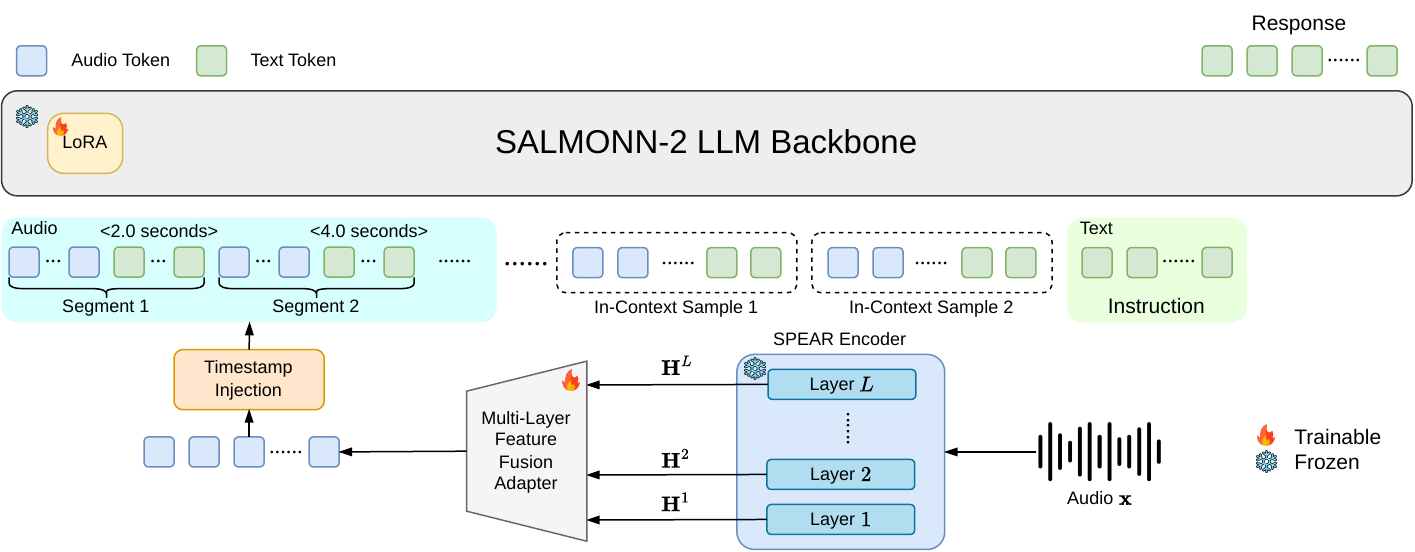}
    \caption{The architecture of SALMONN-2. The unified SSL audio encoder (SPEAR) transforms the input waveform into layer-wise representations, which are aggregated by the MLF adapter. Then, timestamp tokens are injected into audio embeddings to provide explicit temporal grounding cues. Audio-text multimodal entries are further incorporated as contextual cues to equip the LLM with multimodal in-context learning capabilities.}
    \label{fig:model}
\end{figure*}
\vspace{-0.5em}

\section{SALMONN-2: General-Purpose Audio LLM with SSL Representations}

\subsection{Overall Architecture}

SALMONN-2 consists of three main components: a self-supervised audio encoder, an MLF adapter, and a text LLM.
The overall architecture is illustrated in Figure~\ref{fig:model}.
The audio encoder extracts hierarchical representations from all layers.
The MLF adapter aggregates these layer-wise features and maps them into the LLM embedding space, and the text LLM performs task-specific autoregressive generation conditioned on the instruction and optional audio-text context.

To support temporally grounded tasks, timestamp tokens are interleaved with the audio embeddings before being passed to the LLM. SALMONN-2 also supports optional multimodal context, where additional textual or audio-text examples can be provided as part of the conditioning input. This unified formulation allows the same model to handle standard audio understanding, timestamp-aware understanding and multimodal in-context learning tasks.

\subsection{Unified SSL Audio Encoder}

SALMONN-2 adopts SPEAR~\cite{spear} as a unified SSL audio encoder. SPEAR is trained to learn general-purpose representations across speech and audio domains, making it a suitable front-end for ALLMs that require broad speech and audio understanding capabilities.

Given an input waveform $\mathbf{x}$, the encoder produces hidden representations from all transformer layers:
\begin{equation}
\mathcal{H}
=
f_{\mathrm{enc}}(\mathbf{x})
=
\{\mathbf{H}^{1}, \mathbf{H}^{2}, \ldots, \mathbf{H}^{L}\},
\end{equation}
where $L$ denotes the number of encoder layers and $\mathbf{H}^{l}\in\mathbb{R}^{T\times d}$ represents the hidden states from layer $l$, with sequence length $T$ and hidden dimension $d$.

Instead of relying solely on the final-layer representation, SALMONN-2 retains hidden states from all layers, which motivates the multi-layer feature fusion design introduced in the next subsection.

\subsection{Multi-Layer Feature Fusion Adapter}
\label{sec:multi-layer-feature-fusion}

Previous studies have shown that different layers of SSL encoders capture different levels of information~\cite{SslModelEvalTaslp}.
Lower layers tend to encode acoustic and phonetic characteristics, intermediate layers capture speaker-related information, while higher layers contain increasingly semantic representations. Consequently, relying only on the final encoder layer may discard useful information for downstream speech and audio understanding tasks.

To better utilise layer-wise representations, we propose an MLF adapter that performs both multi-layer aggregation and audio-to-text projection. First, layer normalisation is independently applied to the hidden states from each encoder layer, and the normalised representations are concatenated along the feature dimension:
\begin{align}
\widetilde{\mathbf{H}}^{l}
&=
\mathrm{LayerNorm}(\mathbf{H}^{l}), \\
\mathbf{H}_{\mathrm{cat}}
&=
[\widetilde{\mathbf{H}}^{1};
\widetilde{\mathbf{H}}^{2};
\cdots;
\widetilde{\mathbf{H}}^{L}]
\in
\mathbb{R}^{T\times Ld}.
\end{align}

A learnable down-projection is then applied to compress the concatenated representation:
\begin{equation}
\mathbf{Z}
=
\mathbf{H}_{\mathrm{cat}}\mathbf{W}_{d},
\quad
\mathbf{W}_{d}\in\mathbb{R}^{Ld\times d}.
\end{equation}

To reduce the audio token rate and align the features with the LLM input space, the adapter groups every $q$ consecutive frames and concatenates them along the feature dimension:
\begin{equation}
\mathbf{G}_{m}
=
[\mathbf{Z}_{(m-1)q+1};
\ldots;
\mathbf{Z}_{mq}],
\quad
\mathbf{G}_{m}\in\mathbb{R}^{qd}.
\end{equation}

The grouped features are then projected into the LLM embedding space using a trainable projection network:
\begin{equation}
\mathbf{A}_{m}
=
\phi(\mathbf{G}_{m}),
\quad
\mathbf{A}_{m}\in\mathbb{R}^{d_{\mathrm{LLM}}},
\end{equation}
where $\phi(\cdot)$ is implemented as a two-layer MLP. The resulting audio embedding sequence is denoted as $\mathbf{A}=\{\mathbf{A}_{1},\ldots,\mathbf{A}_{M}\}$, where $M=\lfloor T/q \rfloor$.

Compared with weighted-sum aggregation, which collapses all encoder layers using a single set of learnable weights, the proposed MLF adapter preserves layer-wise information before projection. This enables the adapter to learn richer cross-layer interactions and better exploit the hierarchical representations encoded by the SSL model. We empirically compare the proposed MLF adapter with final-layer representations and learnable weighted-sum aggregation in Section~\ref{sec:ablation}.

\subsection{Timestamp Injection}
\label{sec:timestamp-injection}

Temporal grounding is essential for tasks requiring explicit event localisation, such as sound event detection and timestamp-aware audio understanding~\cite{spotsound, shi2026towards, team2026qwen35}. However, the audio embeddings only preserve relative ordering information and do not explicitly encode absolute temporal positions.

Following prior timestamp-aware audio-language models~\cite{team2026qwen35}, we interleave timestamp tokens with audio embeddings before feeding them into the LLM. Let the audio token rate after the MLF adapter be $r$ Hz and the timestamp granularity be $g$ seconds. Timestamp tokens are inserted every $k=r\cdot g$ audio tokens.

For a timestamp corresponding to time $t$, we construct the textual representation \texttt{<t seconds>} and tokenise it using the LLM tokeniser. Rather than introducing additional trainable timestamp embeddings, we directly reuse the corresponding token embeddings from the LLM vocabulary. The resulting timestamp token sequence is inserted between neighbouring audio segments:
\begin{equation}
[
\mathbf{A}_{1:k},
\mathbf{E}_{g},
\mathbf{A}_{k+1:2k},
\mathbf{E}_{2g},
\ldots
],
\end{equation}
where $\mathbf{E}_{t}$ denotes the sequence of token embeddings corresponding to \texttt{<t seconds>}. We denote the timestamp-injected audio sequence as $\widehat{\mathbf{A}}$.

By representing timestamps using existing textual tokens, temporal markers naturally reside in the same semantic space as the language model vocabulary, enabling timestamp prediction and temporal reasoning without introducing additional timestamp-specific parameters.

\subsection{Multimodal In-Context Learning for Contextual ASR}
\label{sec:micl}

Beyond conventional speech and audio understanding tasks, we further investigate whether ALLMs can effectively utilise multimodal contextual information. We study this through contextual ASR, where external context can directly affect the recognition of rare words and named entities.

Conventional contextual ASR typically relies on text-only biasing entries. However, for foreign names, irregular spellings, accented pronunciations, or words with multiple valid pronunciations, the written form alone may not provide sufficient information to associate a biasing entry with its acoustic realisation. To address this limitation, we formulate contextual ASR as a multimodal in-context learning (MICL) task.

Given an input utterance $\mathbf{x}$ and a biasing list containing entries $\{w_i\}_{i=1}^{N_\text{biasing}}$, we construct the multimodal context as
\begin{equation}
\mathcal{C}
=
[\mathbf{s}_{w_1},w_1,\ldots,\mathbf{s}_{w_{N_\text{biasing}}},w_{N_\text{biasing}}],
\end{equation}
where $\mathbf{s}_{w_i}$ denotes a reference pronunciation of word $w_i$.
The complete model input consists of the task instruction, the contextual information $\mathcal{C}$, and the query audio $\mathbf{x}$.
Reference pronunciation audio samples are also encoded by the SPEAR encoder and MLF adapter. Timestamp tokens are not inserted into the reference-pronunciation audio embeddings.

During training, we include both text-only and multimodal contextual ASR examples, enabling SALMONN-2 to support both forms of contextual biasing at inference time.
We also incorporate distractor entries and target-word dropout to mitigate over-biasing, encouraging the model to rely on contextual information only when supported by the input audio.

\subsection{Training Objective}

SALMONN-2 is trained in a unified conditional generation framework. Let $\mathcal{P}$ denote the conditioning input, which may contain a task instruction, timestamp-injected audio embeddings $\widehat{\mathbf{A}}$, and optional multimodal context $\mathcal{C}$. Given a target text sequence $\mathbf{y}=(y_1,\ldots,y_N)$, the model is optimised with the standard autoregressive language modelling objective:
\begin{equation}
\mathcal{L}
=
-\sum_{n=1}^{N}
\log p(y_n \mid y_{<n}, \mathcal{P}).
\end{equation}
The same objective is used across all training tasks, allowing SALMONN-2 to unify conventional audio understanding, timestamp-aware generation and multimodal in-context learning within a single generative framework.
\section{Experimental Setup}

\subsection{Data}

SALMONN-2 is trained with diverse data covering speech, general audio, and music domains. The summary of training datasets for SALMONN-2 is shown in Table~\ref{tab:training_data}.
As can be seen, SALMONN-2 is trained with 18.2k hours of instruction-tuning data in total.
This is substantially smaller than the training scale reported by several recent ALLMs, which often rely on tens of thousands to millions of hours of supervised audio-text paired data~\cite{mimo-audio,audioflamingo3,moss-audio} (see Table~\ref{tab:mmxx}).
SALMONN-2 therefore provides a comparatively data-efficient setting for building general-purpose ALLMs.

\begin{table}[t]
\centering
\caption{Summary of training data used for SALMONN-2.}
\label{tab:training_data}
\resizebox{\linewidth}{!}{
\begin{tabular}{llcc}
\toprule
\textbf{Task} & \textbf{Dataset} & \textbf{\# Samples} & \textbf{\# Hours} \\
\midrule
\multicolumn{4}{c}{\textit{Speech and Paralinguistic Tasks}} \\
\midrule
ASR & LibriSpeech  & 281k & 960 \\
ASR & GigaSpeech   & 910k & 1000 \\
ASR & CommonVoice  & 1.1M & 1740 \\
ER  & IEMOCAP Session 1-4     & 4k & 5.2 \\
ER  & MSP-Podcast  & 129k & 195 \\
SV  & VoxCeleb1    & 523k & 1200 \\
OSR & LibriMix & 65k & 260 \\
\midrule
\multicolumn{4}{c}{\textit{General Audio and Music Tasks}} \\
\midrule
AAC & WavCaps      & 270k & 760 \\
AAC & AudioCaps    & 48k & 134 \\
AAC & Clotho       & 4k & 25 \\
AAC & AudioSet & 2M & 5000 \\
Music captioning & MusicCaps & 2.6k & 7 \\
Music understanding & MillionSong & 48k & 400 \\
Music transcription & MusicNet & 0.3k & 3 \\
\midrule
\multicolumn{4}{c}{\textit{Specialised Tasks and QA Data}} \\
\midrule
Contextual ASR & GigaSpeech & 200k & 220 \\
Contextual ASR & SPGISpeech  & 77k & 196 \\
SQA & QualiSpeech & 84k & 136 \\
SED & AudioSet-strong & 429k & 329 \\
SED & FineLAP100k & 396k & 277 \\
SED & PicoAudio2 & 93k & 172 \\
Spoofing & HoliAntiSpoof & 786k & 1130 \\
General QA & In-house QA data & 1.4M & 4034 \\
\midrule
\multicolumn{2}{l}{\textbf{Total}} & 8.9M & 18183 \\
\bottomrule
\end{tabular}}
\end{table}

In addition to common speech and audio understanding data (e.g., Automatic Speech Recognition (ASR), Automatic Audio Captioning (AAC)), we further incorporate several extended audio analysis data: spoofing analysis, sound event detection (SED), and speech quality assessment (SQA).
For spoofing analysis, we use the training set in HoliAntiSpoof~\cite{xu2026holiantispoof}, which covers both fully spoofed speech and partially spoofed speech generated by diverse spoofing methods.
For SED, we use data from AudioSet-strong~\cite{audioset-strong}, FineLAP~\cite{li2026finelap} and PicoAudio2~\cite{zheng2026picoaudio2}.
We construct two types of training data for SED: question answering (QA) and summarisation.
Here, QA requires the model to determine the occurrence of queried sound events, while summarisation requires the model to describe each sound event together with its corresponding timestamp.
For both spoofing analysis and SED, we reformulate the response into a structured JSON format to facilitate parsing.
For SQA, we use QualiSpeech~\cite{wang2025qualispeech} for training.
With these multifaceted audio analysis datasets, SALMONN-2 acquires more comprehensive audio understanding capabilities. 

\subsection{Implementation Details}

SALMONN-2 adopts a two-stage training pipeline. In the first stage, we mainly use ASR and AAC data to align the audio encoder representation space with the text LLM, together with SED data to introduce timestamp-aware temporal grounding. In the second stage, we incorporate a broader set of task-specific instruction tuning data and audio-text QA data to extend the model to more diverse audio understanding tasks and improve its instruction-following and reasoning abilities. In both stages, the audio encoder is always kept frozen, while the MLF adapter is fully trainable and the text LLM is updated using LoRA.
The training details of the two stages are summarised in Table~\ref{tab:training_details}. 

\begin{table}[t]
\centering
\caption{Training details for the two-stage training pipeline of SALMONN-2.}
\label{tab:training_details}
\begin{tabular}{lccc}
\toprule
Stage & \# Steps & Batch Size & Learning Rate \\
\midrule
Stage 1 & 40k & 192 & 2e-4 \\
Stage 2 & 50k & 192 & 2e-4 \\
\bottomrule
\end{tabular}
\end{table}


\subsection{Model}

We use SPEAR XLarge as the unified SSL audio encoder. It contains 600M parameters and consists of 13 Zipformer~\cite{zipformer} layers, producing 1280-dim frame-level audio representations at 50~Hz.
The MLF adapter described in Section~\ref{sec:multi-layer-feature-fusion} is used to process hidden states of the SPEAR encoder. 
First, encoder hidden states from $L=13$ SPEAR layers are concatenated to $Ld=16640$ dimensions, and then project back into 1280-dim feature.
Then, every $q=5$ consecutive fused frames are concatenated along the feature dimension, reducing the frame rate from 50~Hz to 10~Hz. Finally, the downsampled representations are projected into the LLM embedding space.
The timestamp granularity $g$ in \Cref{sec:timestamp-injection} is set to 2, so $k$ is 20.
The concatenated features are then projected to the LLM embedding dimension with a linear layer, followed by a ReLU activation and another linear layer. This produces a sequence of audio embeddings aligned with the input space of the LLM.


We use Qwen-series LLMs as the text backbone and evaluate SALMONN-2 with different LLM scales.
The detailed model configurations are summarised in Table~\ref{tab:model_config}.

\begin{table}[t]
\centering
\caption{Model configurations used in SALMONN-2. The audio encoder is frozen in all settings, while the LLM is adapted using LoRA and the MLF adapter is fully trainable. The second column denotes the number of parameters in the text LLM.}
\label{tab:model_config}
\adjustbox{width=\linewidth}{
\begin{tabular}{lcccc}
\toprule
Text LLM & \# Params & Embed Dim. & LoRA Rank & \# Params MLF \\
\midrule
Qwen3-8B & 8B & 4096 & 64 & 64M \\
Qwen3-30B-A3B\footnotemark & 30B-A3B & 2048 & 64 & 56M \\
\bottomrule
\end{tabular}}
\end{table}
\footnotetext{\url{https://huggingface.co/Qwen/Qwen3-30B-A3B-Instruct-2507}}

\subsection{MICL-based Contextual ASR}
\label{subsec:micl_contextual_asr_setting}

We construct contextual ASR training data using GigaSpeech~\cite{gigaspeech} and SPGISpeech~\cite{spgispeech}. For each corpus, words outside the 20k most frequent words in the training transcripts are treated as contextual words requiring biasing. Approximately 30\% of the training samples contain at least one contextual word, with an average of 2.2 contextual words per contextual ASR sample. For MICL training, each contextual word is paired with three reference pronunciations generated using OmniVoice~\cite{omnivoice}. These pronunciations are synthesised using different speaker identities drawn from an external pool of 200 TTS-generated speakers, and one pronunciation is randomly selected for each contextual entry during training.

During training, the biasing list contains all ground-truth contextual words together with 25--50 randomly sampled distractor words. For utterances that do not contain contextual words, the biasing list consists entirely of distractors. To reduce over-biasing, each ground-truth contextual word is independently removed from the biasing list with probability 0.1. Furthermore, with probability 0.2, the pronunciation audio is removed and the corresponding example is converted into a text-only contextual biasing sample.

For evaluating the contextual biasing performance, we curate a contextual biasing test set based on GigaSpeech test data, following the same construction protocol as the LibriSpeech biasing benchmark.
Specifically, we use the same biasing vocabulary to identify contextual words in the test samples, and randomly fill the biasing list to lengths of 20, 50 and 100 by sampling distractors from the biasing vocabulary. In the resulting GigaSpeech test set, 37\% of the samples contain at least one biasing word, with an average of 1.5 ground-truth biasing words per contextual ASR sample. For MICL, another pool of TTS-generated speakers is used to synthesise pronunciations of the words.
In addition to standard WER, we report B-WER and U-WER to better analyse contextual biasing behaviour. B-WER is computed on reference words that belong to the ground-truth biasing set, reflecting the model's ability to recognise biased words. U-WER is computed on the remaining reference words, measuring whether contextual biasing degrades recognition of unbiased words. All metrics are reported as percentages.

\subsection{Evaluation Settings}

\begin{table}[ht]
    \centering
    \caption{Tasks, datasets and metrics used for comprehensive evaluation of ALLMs.}
    \adjustbox{width=\linewidth}{
    \begin{tabular}{c|ccc}
    \toprule
    Dimension & Task & Data & Metrics \\
    \midrule
    \multirow{9}{*}{\shortstack{Canonical\\Tasks}} & ASR & LibriSpeech & \%WER \\
     & ASR & GigaSpeech & \%WER \\
     & AAC & AudioCaps & SBERT \\
     & AAC & Clotho & SBERT \\
     & MC & MusicCaps & BLEU4/ROUGE-L \\
     & ER & IEMOCAP Session 5 & Accuracy \\
     & En2Zh & CoVoST2 & BLEU4 \\
     & PR & LibriSpeech & PER \\
     & OSR & LibriMix & \%cpWER \\
     & SV & VoxCeleb1 & Accuracy \\
    \midrule
     \multirow{3}{*}{\shortstack{General Audio\\Understanding\\Benchmarks}} & QA & MMAU-Pro & Accuracy \\
     & QA & MMAR & Accuracy \\
     & QA & MMSU & Accuracy \\
    \midrule
    \multirow{2}{*}{\shortstack{MICL\\Contextual ASR}} & TB-ASR & GigaSpeech & \%WER \\
     & MB-ASR & GigaSpeech & \%WER \\
    \midrule
    \multirow{3}{*}{\shortstack{Extended Audio\\Analysis Tasks}} & SED & DESED & mIOU \\
     & Spoofing & PartialEdit & mIOU \\
     & SQA & QualiSpeech & PCC \\
    \bottomrule
    \end{tabular}
    }
    \label{tab:eval_settings}
\end{table}

We evaluate SALMONN-2 from four dimensions to comprehensively assess its general-purpose audio understanding capabilities, as summarised in \Cref{tab:eval_settings}.
First, following the SALMONN~\cite{salmonn} evaluation protocol, we evaluate the model on \textit{canonical tasks}, including ASR, AAC, music captioning (MC), emotion recognition (ER), speech translation (En2Zh), overlapped speech recognition (OSR), and speaker verification (SV), covering fundamental audio understanding capabilities.
The only difference to the SALMONN evaluation protocol is that for AAC, we use SentenceBERT (SBERT)~\cite{sentencebert} similarity instead of n-gram-based metrics, to minimise the influence of stylistic variations in generated captions.
We then evaluate SALMONN-2 on representative \textit{ALLM understanding benchmarks}, including MMAU-Pro~\cite{mmau-pro}, MMAR~\cite{mmar}, and MMSU~\cite{mmsu}, to assess its audio understanding and reasoning abilities on challenging QA tasks.
We further investigate the MICL capability of SALMONN-2 on contextualised ASR, as described in \Cref{subsec:micl_contextual_asr_setting}.
Two evaluation settings are considered: text-biasing ASR (TB-ASR), where only textual biasing words are provided as context, and multimodal-biasing ASR (MB-ASR), where both textual biasing words and their pronunciations are provided.
Finally, the evaluation is conducted on the extended audio analysis tasks, including SED, spoofing analysis, and SQA.
DESED~\cite{desed}, PartialEdit~\cite{partialedit}, and QualiSpeech are adopted as the evaluation datasets for SED, spoofing analysis, and SQA, respectively.
Since both SED and spoofing analysis require the model to predict the temporal spans of queried sound events or spoofed regions, we use mean intersection over union (mIOU) as the metric.
For SQA, we use the Pearson correlation coefficient (PCC) to measure the correlation between predicted and ground-truth quality scores.
\section{Experimental Results}\label{sec: experimental results}

\subsection{Canonical Tasks}

\begin{table*}[ht]
    \centering
    \caption{Performance comparison of open-weight ALLMs on canonical audio understanding tasks. WER on LibriSpeech test-clean and test-other are listed and separated using ``/". ``AF-3" and ``AF-Next" stand for AudioFlamingo3 (AF3) and AudioFlamingo-Next (AF-Next), respectively. We report the official results from the corresponding papers or technical reports whenever available; otherwise, we evaluate the models using heuristically optimised prompts.
    $^\ast$ indicates that the low score is mainly due to the model failing to follow task instructions. Best and second-best results are highlighted in bold and underlined, respectively.}
    \adjustbox{width=\linewidth}{
    \begin{tabular}{cc|ccccccccc}
    \toprule
    Model & LLM Size & LibriSpeech$\downarrow$ & GigaSpeech$\downarrow$ & AudioCaps$\uparrow$ & Clotho$\uparrow$ & MusicCaps$\uparrow$ & IEMOCAP$\uparrow$ & CoVoST2$\uparrow$ & LibriMix$\downarrow$ & VoxCeleb1$\uparrow$ \\
    \midrule
    SALMONN~\cite{salmonn} & 7B & 2.1/4.9 & 10.0 & 57.6 & 44.7 & 5.5/21.8 & 69.0 & 33.1 & 23.0 & \underline{94.0} \\
    Qwen2.5-Omni~\cite{qwen2.5-omni} & 8B & 1.8/3.4 & 23.0 & 64.4 & 50.2 & 5.5/\underline{22.5} & 63.5 & 41.4 & 55.3 & 62.9 \\
    Kimi-Audio~\cite{kimi-audio} & 8B & \textbf{1.4/2.6} & 9.1 & 52.0 & 50.7 & 0.0/9.0 & \textbf{76.2} & 0.3$^\ast$ & 44.4 & 61.9 \\
    MiMo-Audio~\cite{mimo-audio} & 8B & 3.5/6.4 & 14.1 & 53.6 & 44.2 & 0.8/15.9 & 61.2 & 35.8 & 57.7 & 50.0 \\
    AF-3~\cite{audioflamingo3} & 8B & 1.6/3.1 & 9.3 & 64.0 & \underline{54.4} & 4.0/19.1 & 63.8 & 11.9$^\ast$ & 52.5 & 50.0 \\
    AF-Next~\cite{audioflamingo-next} & 9B & \underline{1.5/2.8} & 9.2 & 61.1 & \textbf{54.9} & 3.1/19.2 & 55.0 & 1.7$^\ast$ & 67.6 & 50.0 \\
    MOSS-Audio~\cite{moss-audio} & 9B & 2.2/5.6 & \textbf{8.6} & 47.4 & 44.2 & 3.7/19.6 & 62.7 & 23.9 & 45.9 & 61.0 \\
    Ours, SALMONN-2 & 9B & 1.7/3.0 & 9.0 & \underline{65.0} & 53.4 & \textbf{5.9}/22.0 & 71.0 & \textbf{46.6} & \textbf{8.8} & 93.8 \\
    \midrule
    Ours, SALMONN-2 & 30B-A3B & 1.6/3.0 & \underline{8.9} & \textbf{65.2} & 53.6 & \underline{5.6}/\textbf{23.5} & \underline{74.0} & \underline{46.4} & \underline{9.0} & \textbf{94.1} \\
    \bottomrule
    \end{tabular}
    }
    \label{tab:canonical_tasks}
\end{table*}

\Cref{tab:canonical_tasks} compares SALMONN-2 with representative open-weight ALLMs on canonical audio understanding tasks.
We compare the 9B-version SALMONN-2 with other ALLMs of a similar scale.
SALMONN-2 demonstrates consistently strong performance across diverse domains.
On ASR, SALMONN-2 achieves competitive WERs on both LibriSpeech and GigaSpeech.
For AAC, SALMONN-2 achieves the best performance on AudioCaps and competitive performance on Clotho.
Similarly, for MC, SALMONN-2 achieves the best BLEU score and the second-best ROUGE-L score.
In contrast, several existing ALLMs exhibit imbalanced performance across tasks and datasets.
For example, Kimi-Audio achieves the best speech recognition performance on LibriSpeech but performs less favourably on audio and music captioning.
Qwen2.5-Omni and MiMo-Audio perform well on LibriSpeech but their WERs increase notably on GigaSpeech.
Moreover, we observe that MiMo-Audio occasionally transcribes English utterances into Chinese on GigaSpeech, while Qwen2.5-Omni sometimes refuses to transcribe the given speech. 
These failure cases lead to high WERs for the two models, indicating potential instability when generalising across different datasets.

Beyond recognition and captioning, SALMONN-2 also demonstrates competitive performance on other understanding tasks.
SALMONN-2 achieves the second-highest accuracy on IEMOCAP emotion recognition and the best performance on CoVoST2 speech translation.
AudioFlamingo3 (AF3), AudioFlamingo-Next (AF-Next), and Kimi-Audio obtain low BLEU scores because they do not follow the instruction (doing translation), outputting speech transcriptions instead of translation results.
For OSR and SV, most existing ALLMs fail to perform the tasks reliably, possibly because these tasks are rarely covered by their instruction-tuning data.
These results indicate that although current ALLMs achieve satisfactory performance on standard audio understanding tasks, they still struggle to generalise to tasks that are not explicitly seen during training.
Compared with SALMONN, which was explicitly trained on OSR and SV, SALMONN-2 significantly reduces the cpWER on LibriMix while maintaining comparable performance on VoxCeleb1.

Overall, benefiting from the improved architecture and training data, SALMONN-2 substantially outperforms SALMONN.
Compared with existing open-weight ALLMs, SALMONN-2 consistently ranks among the top-performing models across all canonical tasks, demonstrating balanced and robust audio understanding capabilities across diverse domains and tasks.
Further scaling the model size to 30B does not yield substantial improvements.
This may indicate that these canonical tasks are not challenging enough, or that they rely more heavily on audio perception than on advanced textual understanding and reasoning.

\subsection{General Audio Understanding Benchmarks}

\begin{table}[t]
\centering
\caption{Performance comparison of similar-sized ALLMs on speech and audio understanding benchmarks. 
``Data'' denotes the reported amount of supervised audio-text paired data used for ALLM training (in hours). 
Values are approximate and may not be strictly comparable due to differences in data definitions.}
\label{tab:mmxx}
\adjustbox{max width=\linewidth}{
\begin{tabular}{lccccc}
\toprule
Model & LLM Size & Data (h) & MMAU-Pro & MMAR & MMSU \\
\midrule
Qwen2.5-Omni~\cite{qwen2.5-omni}        & 8B & --    & 52.2 & 56.7 & 61.3 \\
Kimi-Audio~\cite{kimi-audio}          & 8B & $>$13M       & 56.6 & 60.8 & 54.7 \\
MiMo-Audio~\cite{mimo-audio}          & 8B & $>$1M  & 53.4 & 61.7 & 61.9 \\
AF-3~\cite{audioflamingo3}      & 8B & $>$55k & 51.7 & 58.5 & 61.4 \\
AF-Next~\cite{audioflamingo-next}  & 9B & $>$500k       & 56.3 & 59.7 & 59.4 \\
MOSS-Audio~\cite{moss-audio}          & 9B & $>$1M  & 57.5 & 64.4 & 66.4 \\
Ours, SALMONN-2           & 9B & 18.2k  & \textbf{58.5} & \textbf{64.5} & \textbf{69.5} \\
\midrule
Ours, SALMONN-2 & 30B-A3B & 18.2k & 60.3 & 67.6 & 72.0 \\
\bottomrule
\end{tabular}
}
\end{table}

Table~\ref{tab:mmxx} compares SALMONN-2 with recent open-weight ALLMs on three general audio understanding benchmarks. SALMONN-2 achieves the best performance across all evaluated benchmarks, obtaining 58.5 on MMAU-Pro, 64.5 on MMAR and 69.5 on MMSU. Compared with the strongest baseline, MOSS-Audio, SALMONN-2 improves the scores by 1.0, 0.1 and 3.1 points on MMAU-Pro, MMAR and MMSU, respectively. The improvement is particularly clear on MMSU, suggesting stronger performance on speech-centric semantic and paralinguistic tasks.

Notably, SALMONN-2 achieves these results while using substantially less supervised training data than several recent ALLMs. This demonstrates the strong data efficiency of SALMONN-2, and suggests that using a unified SSL audio encoder in ALLMs is a promising direction for reducing the reliance on large-scale supervised audio-language data.

Compared with canonical tasks, scaling the model size yields substantial performance gains on general audio understanding benchmarks.
The 30B model achieves an average improvement of 2.5 points across the three benchmarks, indicating that these benchmarks place greater demands on the language understanding and reasoning capabilities of LLMs.

\subsection{MICL Contextual ASR}

\begin{table*}[t]
\centering
\caption{Contextual ASR results on GigaSpeech. Each entry reports WER(\%)/B-WER(\%)/U-WER(\%) (defined in Section~\ref{subsec:micl_contextual_asr_setting}).}
\label{tab:contextual_asr_gigaspeech}
\begin{tabular}{lllcccc}
\toprule
\multirow{2}{*}{\textbf{Model}} & \multirow{2}{*}{\textbf{MICL training}} & \multirow{2}{*}{\textbf{Biasing Method}} & \multicolumn{4}{c}{\textbf{GigaSpeech}} \\
\cmidrule(lr){4-7}
& & & No-biasing & $L=20$ & $L=50$ & $L=100$ \\
\midrule
AF-3~\cite{audioflamingo3} & $\times$ & Text & 9.3 / 15.8 / 9.3 & 18.0 / 19.0 / 17.9 & 17.9 / 19.7 / 17.7 & 19.8 / 21.0 / 19.7 \\
\midrule
\multirow{2}{*}{MOSS-Audio~\cite{moss-audio}} & \multirow{2}{*}{$\times$} & Text & \multirow{2}{*}{8.6 / 15.2 / 8.4} & 8.7 / 9.1 / 8.6 & 8.9 / 9.6 / 8.8 & 9.3 / 10.2 / 9.3 \\
& &  MICL &  & 26.5 / 26.0 / 26.6 & 53.9 / 31.8 / 54.6 & 65.5 / 36.9 / 66.4 \\
\midrule
\multirow{4}{*}{SALMONN-2} & \multirow{2}{*}{$\times$} & Text & \multirow{2}{*}{9.1 / 15.5 / 9.0} & 9.1 / 12.2 / 9.0 & 9.2 / 14.1 / 9.0 & 9.2 / 15.6 / 8.8 \\
&  & MICL & & 25.7 / 26.0 / 25.5 & 47.7 / 29.7 / 50.2 &  57.3 / 33.5 / 58.1 \\
\cmidrule(lr){2-7}
& \multirow{2}{*}{$\checkmark$} & Text & \multirow{2}{*}{9.0 / 15.4 / 8.9} & 8.3 / 7.5 / 8.3 & 8.5 / 7.9 / 8.4 & 8.7 / 8.8 / 8.6 \\
& & MICL &  & 8.2 / 6.8 / 8.3 & 8.4 / 7.2 / 8.4 & 8.5 / 8.0 / 8.5 \\

\bottomrule
\end{tabular}
\end{table*}

Table~\ref{tab:contextual_asr_gigaspeech} presents the contextual ASR results on GigaSpeech. The biasing method indicates the type of contextual information provided to the model: \textit{Text} uses only textual biasing entries, while \textit{MICL} provides both the textual entry and a reference pronunciation as multimodal context. We evaluate different biasing list lengths $L$, where a larger $L$ introduces more distractor entries and therefore requires stronger contextual selection ability. 

For existing ALLMs, contextual ASR remains challenging without task-specific training.
AF-3 shows substantial degradation when textual biasing entries are provided, with WER increasing from 9.3 without biasing to 19.8 with $L=100$. 
Although MOSS-Audio is more robust with text-only biasing, its MICL performance deteriorates rapidly as the biasing list grows, reaching 65.5 WER when $L=100$.
Similarly, the variant of SALMONN-2 trained without MICL also struggles to leverage contextual information, exhibiting substantial performance degradation when presented with a biasing list.
These results suggest that existing ALLMs do not naturally learn to utilise multimodal contextual information for ASR, especially when the context contains many distractors.

In contrast, SALMONN-2 effectively benefits from both textual and multimodal contextual information. With text-only biasing, SALMONN-2 consistently improves over the no-biasing baseline and achieves lower B-WER across all list lengths. Using MICL further improves contextual word recognition, reducing B-WER from 7.5 to 6.8 at $L=20$ and from 8.8 to 8.0 at $L=100$. This indicates that the pronunciation examples provide useful acoustic contextual information, enabling more targeted biasing than text-only context. Moreover, MICL maintains strong performance as the biasing list becomes longer, suggesting better robustness to distractors while preserving stable U-WER on non-biased words. Overall, these results show that SALMONN-2 can acquire effective multimodal in-context learning for contextual ASR through targeted training.

\subsection{Extended Audio Analysis Tasks}

\begin{table}[ht]
    \centering
    \caption{Performance comparison on extended audio analysis tasks. "QS.-FT-SALMONN" denotes SALMONN finetuned on the QualiSpeech dataset in \cite{wang2025qualispeech}.}
    \begin{tabular}{l|ccc}
    \toprule
    Model & DESED & Spoofing & QualiSpeech \\
    \midrule
    \multicolumn{4}{c}{\textbf{\textit{Specialised Models}}} \\
    \midrule
    SpotSound-Q~\cite{spotsound} & 61.1 & - & - \\
    HoliAntiSpoof~\cite{xu2026holiantispoof} & - & 71.58 & - \\
    QS.-FT-SALMONN~\cite{wang2025qualispeech} & - & - & 0.660 \\
    \midrule
    \multicolumn{4}{c}{\textbf{\textit{General-Purpose Audio LLMs}}} \\
    \midrule
    SALMONN-2 & \textbf{70.15} & \textbf{73.65} & \textbf{0.661} \\
    MiMo-Audio & 27.65 & 14.27 & 0.358 \\
    MOSS-Audio & 48.14 & 19.14 & 0.402 \\
    \bottomrule
    \end{tabular}
    \label{tab:extended_audio_analysis_tasks}
\end{table}

We further evaluate SALMONN-2 on the extended audio analysis tasks.
Besides recent ALLMs, we also include models specialised for each task for comparison.
These models are also based on the LLM architecture and trained on similar data.
As shown in Table~\ref{tab:extended_audio_analysis_tasks}, SALMONN-2 shows significant advantages over existing ALLMs on SED, benefiting from the timestamp injection mechanism and timestamp-aware QA and summarisation data.
With large-scale training across diverse domains and tasks, SALMONN-2 also surpasses the specialised temporal grounding ALLM SpotSound~\cite{spotsound}, which also employs the timestamp injection mechanism.
For spoofing analysis and SQA, SALMONN-2 substantially outperforms baseline ALLMs and achieves performance comparable to or better than specialised competitors.
These results suggest that SALMONN-2 can serve as a comprehensive audio analysis model, capable of accurately understanding both semantic audio content and low-level acoustic characteristics.




\subsection{Ablation Studies}
\label{sec:ablation}

\begin{table*}[ht]
\centering
\caption{Comparison of different encoder architectures and feature fusion strategies on standard speech, audio, music and paralinguistic tasks. ``LS." denotes LibriSpeech. For LS. (ASR), WERs on the clean/other test sets are reported. For LS. (PR), PER on the clean test set is reported.}
\label{tab:encoder_ablation}
\adjustbox{width=\linewidth}{
\begin{tabular}{lccccccccc}
\toprule
Encoder Config. & AudioCaps$\uparrow$ & MusicCaps$\uparrow$ & LS. (ASR)$\downarrow$ & GigaSpeech$\downarrow$ & CoVoST2$\uparrow$ & LS. (PR)$\downarrow$ & LibriMix$\downarrow$ & IEMOCAP$\uparrow$ & VoxCeleb1$\uparrow$ \\
\midrule

\multicolumn{10}{c}{\textit{\textbf{Dual-Encoder}}} \\
\midrule
Whisper + BEATs & 
\textbf{68.3} & 5.9/22.7 & 2.1/4.9 & \textbf{9.6} & 43.4 & 3.1 & 14.8 & 69.1 & 93.2 \\
WavLM + Dasheng & 
59.5 & 4.9/21.8 & 2.3/4.3 & 11.2 & 39.5 & 3.5 & 43.2 & 56.3 & 81.0 \\
\midrule

\multicolumn{10}{c}{\textit{\textbf{Single Supervised Encoder}}} \\
\midrule
Qwen2.5-Omni Encoder & 
63.6 & 5.4/21.8 & 2.6/6.2 & 11.7 & 39.5 & 9.0 & 47.2 & 63.1 & 88.6 \\
AF-Whisper & 
67.2 & \textbf{10.2}/\textbf{28.8} & 2.0/4.2 & 12.5 & 44.8 & 3.4 & 25.8 & \textbf{73.0} & 86.9 \\
\midrule

\multicolumn{10}{c}{\textit{\textbf{Unified SSL Encoder (SPEAR)}}} \\
\midrule
Last Layer & 
65.6 & 5.5/22.4 & 1.7/3.3 & \textbf{9.6} & 45.9 & 2.9 & 10.0 & 69.8 & 91.9 \\
Weighted-Sum & 
65.7 & 5.6/22.6 & 1.8/3.3 & 9.8 & 45.8 & 2.8 & 12.0 & 69.9 & 93.2 \\
MLF & 
68.1 & 5.8/22.6 & \textbf{1.6}/\textbf{3.2} & \textbf{9.6} & \textbf{45.9} & \textbf{2.7} & \textbf{9.7} & 70.5 & \textbf{94.7} \\
\bottomrule
\end{tabular}
}
\end{table*}

\subsubsection{Effect of Audio Encoder and Feature Fusion}
In this group of ablations, we investigate the following two questions: 
(i) whether a unified SSL encoder is a better foundation for SALMONN-2 than dual encoders or supervised single encoders;
(ii) how encoder representations should be aggregated before being passed to the LLM. 

To ensure a controlled comparison, all ablation models are trained only on a subset of the data in Table~\ref{tab:training_data} by excluding SPGISpeech, MSP-Podcast and all data from the last section in Table~\ref{tab:training_data} (specialised tasks and QA). For all ablation models, the audio encoder is kept frozen, while the MLF adapter and LoRA parameters of the LLM are updated.
All models are trained for 30k steps for faster validation under the same optimisation setting.
For dual-encoder baselines, we follow the strategy used in SALMONN~\cite{salmonn} by concatenating the representations from two encoders along the feature dimension before projecting them into the LLM embedding space.

Table~\ref{tab:encoder_ablation} summarises the results. We first compare different encoder architectures, including dual-encoder systems, supervised single encoders and the proposed SSL encoder. Despite using only a single encoder, SPEAR achieves performance comparable to the dual-encoder Whisper + BEATs baseline, and performs better on several speech-centric tasks such as ASR, PR and OSR. Compared with supervised single encoders, SPEAR also exhibits a more balanced performance profile across diverse task categories. For example, AF-Whisper performs strongly on MC and ER, but is substantially weaker on other tasks including ASR-Giga, OSR and SV.
These results suggest that general-purpose SSL representations are well suited for building unified ALLMs, as they can capture transferable information across speech, audio, music and paralinguistic domains.

We further compare different ways of utilising SPEAR representations. 
MLF consistently outperforms the final-layer baseline, whereas weighted-sum aggregation brings only partial gains, suggesting that intermediate SSL representations are beneficial but require an effective aggregation mechanism to be fully exploited.
Overall, MLF achieves the best performance profile across the evaluated tasks. We hypothesise that although weighted-sum aggregation can combine information from different layers, it learns a single set of layer weights shared across all downstream tasks, which may struggle to balance tasks requiring different types of information. In contrast, MLF preserves layer-wise information before projection and provides stronger representational capacity, enabling the model to better utilise the hierarchical representations learned by the SSL encoder.

Overall, these ablation results show that the unified SSL encoder and MLF form a coherent design: SPEAR provides broad and transferable representations across tasks, while MLF is specifically designed to expose and combine the multi-level information encoded throughout the SSL hierarchy.



\subsubsection{Effect of Timestamp Injection}

We further ablate the timestamp injection module introduced in Section~\ref{sec:timestamp-injection}. Specifically, we compare the standard SALMONN-2 model with a variant trained without timestamp injection. The comparison is conducted on SED and three general benchmarks, namely MMAU-Pro, MMAR and MMSU.
The results are shown in Table~\ref{tab:timestamp_ablation}.

As can be seen, removing timestamp injection leads to a clear drop on SED from 70.15 to 68.74. This shows that interleaving timestamp embeddings with audio embeddings provides useful temporal information for event localisation. Importantly, this improvement is achieved without expanding the LLM vocabulary or introducing additional timestamp-specific embedding parameters.

The benefit of timestamp injection is not limited to the dedicated SED task. SALMONN-2 with timestamp injection also consistently improves over the variant without timestamps on all three general audio-language benchmarks, suggesting that explicit temporal grounding can also benefit broader audio understanding tasks.
Overall, these results validate the effectiveness of timestamp injection as a simple and parameter-efficient mechanism for enhancing temporally grounded audio-language modelling.

\begin{table}[]
\centering
\caption{Effect of timestamp injection on SED and three general audio-language benchmarks.}
\label{tab:timestamp_ablation}
\begin{tabular}{lcccc}
\toprule
Model & SED & MMAU-Pro & MMAR & MMSU \\
\midrule
SALMONN-2 & 70.15 & 58.5 & 64.5 & 69.5 \\
\quad -- w/o timestamp & 68.74  & 58.2 & 64.3  & 69.3 \\
\bottomrule
\end{tabular}
\end{table}

\section{Conclusion}

In this work, we present SALMONN-2, a substantial extension of SALMONN with self-supervised representations.
We conduct a systematic study of audio perception, audio-language adaptation, and multimodal contextual learning.
Specifically, SALMONN-2 adopts SPEAR, a single general-purpose SSL audio encoder, as the unified audio encoder, and introduces the MLF adapter to effectively exploit the hierarchical audio representations.
We further incorporate timestamp injection for temporally grounded audio understanding and explicit MICL training for contextual ASR, enabling the model to utilise both textual and acoustic contextual information.
Extensive experiments show that SALMONN-2 achieves state-of-the-art performance on general audio understanding benchmarks, including MMAU-Pro, MMAR, and MMSU, while using only a 9B-scale LLM and less than 20K hours of instruction-tuning data.
It also demonstrates strong general-purpose audio understanding capabilities on extended audio analysis tasks, such as SED, spoofing analysis, and SQA.
Ablation studies further validate that the unified SSL encoder and the proposed MLF adapter form a simple yet effective alternative to more complex multi-encoder architectures.
Overall, these results demonstrate that SALMONN-2 achieves strong, balanced and data-efficient audio understanding capabilities through effective utilisation of self-supervised audio representations.




\bibliographystyle{IEEETran}
\bibliography{refs}

@article{SslModelEvalTaslp,
  author={Yang, Shu-wen and Chang, Heng-Jui and Huang, Zili and Liu, Andy T. and Lai, Cheng-I and Wu, Haibin and Shi, Jiatong and Chang, Xuankai and Tsai, Hsiang-Sheng and Huang, Wen-Chin and Feng, Tzu-hsun and Chi, Po-Han and Lin, Yist Y. and Chuang, Yung-Sung and Huang, Tzu-Hsien and Tseng, Wei-Cheng and Lakhotia, Kushal and Li, Shang-Wen and Mohamed, Abdelrahman and Watanabe, Shinji and Lee, Hung-yi},
  journal={IEEE/ACM Transactions on Audio, Speech, and Language Processing}, 
  title={{A Large-Scale Evaluation of Speech Foundation Models}}, 
  year={2024},
  volume={32},
  pages={2884-2899},
}

@inproceedings{beats,
  title={{BEATs: Audio Pre-Training with Acoustic Tokenizers}},
  author={Chen, Sanyuan and Wu, Yu and Wang, Chengyi and Liu, Shujie and Tompkins, Daniel and Chen, Zhuo and Wei, Furu},
  booktitle={Proc. ICML},
  year={2023},
  address={Hawaii},
}

@article{chen2022wavlm,
  title={{WavLM: Large-Scale Self-Supervised Pre-Training for Full Stack Speech Processing}},
  author={Chen, Sanyuan and Wang, Chengyi and Chen, Zhengyang and Wu, Yu and Liu, Shujie and Chen, Zhuo and Li, Jinyu and Kanda, Naoyuki and Yoshioka, Takuya and Xiao, Xiong and others},
  journal={IEEE Journal of Selected Topics in Signal Processing},
  volume={16},
  number={6},
  pages={1505--1518},
  year={2022},
  publisher={IEEE},
}

@inproceedings{spear,
  title={{SPEAR: A Unified SSL Framework for Learning Speech and Audio Representations}},
  author={Yang, Xiaoyu and Yang, Yifan and Jin, Zengrui and Cui, Ziyun and Wu, Wen and Li, Baoxiang and Zhang, Chao and Woodland, Phil},
  booktitle={Proc. ICML},
  year={2026},
  address={Seoul},
}

@inproceedings{salmonn,
  title={{SALMONN: Towards Generic Hearing Abilities for Large Language Models}},
  author={Tang, Changli and Yu, Wenyi and Sun, Guangzhi and Chen, Xianzhao and Tan, Tian and Li, Wei and Lu, Lu and MA, Zejun and Zhang, Chao},
  booktitle={Proc. ICLR},
  year={2024},
  address={Vienna}
}

@article{qwen2.5-omni,
  title={{Qwen2.5-Omni Technical Report}},
  author={Xu, Jin and Guo, Zhifang and He, Jinzheng and Hu, Hangrui and He, Ting and Bai, Shuai and Chen, Keqin and Wang, Jialin and Fan, Yang and Dang, Kai and others},
  journal={arXiv preprint arXiv:2503.20215},
  year={2025}
}

@inproceedings{gigaspeech,
  title={{GigaSpeech: An Evolving, Multi-domain {ASR} Corpus with 10,000 Hours of Transcribed Audio}},
  author={Chen, Guoguo and Chai, Shuzhou and Wang, Guanbo and Du, Jiayu and others},
  booktitle={Proc. Interspeech},
  year={2021},
  address={Brno},
}

@article{mimo-audio,
  title={{MiMo-Audio: Audio Language Models are Few-Shot Learners}},
  author={Zhang, Dong and Wang, Gang and Xue, Jinlong and Fang, Kai and Zhao, Liang and Ma, Rui and Ren, Shuhuai and Liu, Shuo and Guo, Tao and Zhuang, Weiji and others},
  journal={arXiv preprint arXiv:2512.23808},
  year={2025}
}

@inproceedings{spgispeech,
  title={{SPGISpeech: 5,000 Hours of Transcribed Financial Audio for Fully Formatted End-to-End Speech Recognition}},
  author={O’Neill, Patrick K and Lavrukhin, Vitaly and Majumdar, Somshubra and Noroozi, Vahid and Zhang, Yuekai and Kuchaiev, Oleksii and Balam, Jagadeesh and Dovzhenko, Yuliya and Freyberg, Keenan and Shulman, Michael D and others},
  booktitle={Proc. Interspeech},
  year={2021},
  address={Brno}
}

@inproceedings{zipformer,
  title={{ZipFormer: A Faster and Better Encoder for Automatic Speech Recognition}},
  author={Yao, Zengwei and Guo, Liyong and Yang, Xiaoyu and Kang, Wei and Kuang, Fangjun and Yang, Yifan and Jin, Zengrui and Lin, Long and Povey, Daniel},
  booktitle={Proc. ICLR},
  year={2024},
  address={Vienna},
}

@article{team2026qwen35,
  title={{Qwen3.5-Omni Technical Report}},
  author={{Qwen Team}},
  journal={arXiv preprint arXiv:2604.15804},
  year={2026}
}

@article{qwen2audio,
  title={{Qwen2-Audio Technical Report}},
  author={Chu, Yunfei and Xu, Jin and Yang, Qian and Wei, Haojie and Wei, Xipin and Guo, Zhifang and Leng, Yichong and Lv, Yuanjun and He, Jinzheng and Lin, Junyang and others},
  journal={arXiv preprint arXiv:2407.10759},
  year={2024}
}

@article{step-audio-r1.5,
  title={{Step-Audio-R1.5 Technical Report}},
  author={Zhang, Yuxin and Zhang, Xiangyu Tony and Liu, Daijiao and Tian, Fei and Deng, Yayue and Chen, Jun and Lin, Qingjian and Zhang, Haoyang and Li, Yuxin and Gong, Jinglan and others},
  journal={arXiv preprint arXiv:2604.25719},
  year={2026}
}

@article{kimi-audio,
  title={{Kimi-Audio Technical Report}},
  author={Ding, Ding and Ju, Zeqian and Leng, Yichong and Liu, Songxiang and Liu, Tong and Shang, Zeyu and Shen, Kai and Song, Wei and Tan, Xu and Tang, Heyi and others},
  journal={arXiv preprint arXiv:2504.18425},
  year={2025}
}

@inproceedings{omnivoice,
  title={{OmniVoice: Towards Omnilingual Zero-Shot Text-to-Speech with Diffusion Language Models}},
  author={Zhu, Han and Ye, Lingxuan and Kang, Wei and Yao, Zengwei and Guo, Liyong and Kuang, Fangjun and Han, Zhifeng and Zhuang, Weiji and Lin, Long and Povey, Daniel},
  booktitle={Proc. Interspeech},
  year={2026},
  address={Sydney}
}

@article{spotsound,
  title={{SpotSound: Enhancing Large Audio-Language Models with Fine-Grained Temporal Grounding}},
  author={Sun, Luoyi and Zhou, Xiao and Li, Zeqian and Zhang, Ya and Wang, Yanfeng and Xie, Weidi},
  journal={arXiv preprint arXiv:2604.13023},
  year={2026}
}

@inproceedings{wang2025qualispeech,
  title={{Qualispeech: A Speech Quality Assessment Dataset with Natural Language Reasoning and Descriptions}},
  author={Wang, Siyin and Yu, Wenyi and Chen, Xianzhao and Tian, Xiaohai and Zhang, Jun and Lu, Lu and Tsao, Yu and Yamagishi, Junichi and Wang, Yuxuan and Zhang, Chao},
  booktitle={Proc. ACL},
  year={2025},
  address={Vienna}
}

@article{xu2026holiantispoof,
  title={{HoliAntiSpoof: Audio LLM for Holistic Speech Anti-Spoofing}},
  author={Xu, Xuenan and Ren, Yiming and Liu, Liwei and Wu, Wen and Li, Baoxiang and Lu, Chaochao and Wang, Shuai and Zhang, Chao},
  journal={arXiv preprint arXiv:2602.04535},
  year={2026}
}

@inproceedings{audioset-strong,
  title={{The Benefit of Temporally-Strong Labels in Audio Event Classification}},
  author={Hershey, Shawn and Ellis, Daniel PW and Fonseca, Eduardo and Jansen, Aren and Liu, Caroline and Moore, R Channing and Plakal, Manoj},
  booktitle={Proc. ICASSP},
  year={2021},
  address={Toronto}
  
}

@inproceedings{li2026finelap,
  title={{FineLAP: Taming Heterogeneous Supervision for Fine-grained Language-Audio Pretraining}},
  author={Li, Xiquan and Xu, Xuenan and Ma, Ziyang and Chen, Wenxi and He, Haolin and Kong, Qiuqiang and Chen, Xie},
  booktitle={Proc. ACL},
  year={2026},
  address={San Diego}
}

@inproceedings{zheng2026picoaudio2,
  title={{PicoAudio2: Temporal Controllable Text-to-Audio Generation with Natural Language Description}},
  author={Zheng, Zihao and Xie, Zeyu and Xu, Xuenan and Wu, Wen and Zhang, Chao and Wu, Mengyue},
  booktitle={Proc. ICASSP},
  year={2026},
  address={Barcelona}
}

@article{qwenaudio,
  title={{Qwen-Audio: Advancing Universal Audio Understanding via Unified Large-Scale Audio-Language Models}},
  author={Chu, Yunfei and Xu, Jin and Zhou, Xiaohuan and Yang, Qian and Zhang, Shiliang and Yan, Zhijie and Zhou, Chang and Zhou, Jingren},
  journal={arXiv preprint arXiv:2311.07919},
  year={2023}
}

@inproceedings{audioflamingo3,
  title={{Audio Flamingo 3: Advancing Audio Intelligence with Fully Open Large Audio Language Models}},
  author={Ghosh, Sreyan and Goel, Arushi and Kim, Jaehyeon and Kumar, Sonal and Kong, Zhifeng and Lee, Sang-gil and Yang, Chao-Han and Duraiswami, Ramani and Manocha, Dinesh and Valle, Rafael and others},
  booktitle={Proc. NeurIPS},
  year={2026},
  address={San Diego}
}

@article{moss-audio,
  title={{MOSS-Audio Technical Report}},
  author={Yang, Chen and Yu, Chufan and Chen, Hanfu and Zhu, Jie and Chen, Jingqi and Chen, Ke and Wang, Wenxuan and Wang, Yang and Jiang, Yaozhou and Jiang, Yi and others},
  journal={arXiv preprint arXiv:2606.01802},
  year={2026}
}

@article{midashenglm,
  title={{MidashengLM: Efficient Audio Understanding with General Audio Captions}},
  author={Dinkel, Heinrich and Li, Gang and Liu, Jizhong and Luan, Jian and Niu, Yadong and Sun, Xingwei and Wang, Tianzi and Xiao, Qiyang and Zhang, Junbo and Zhou, Jiahao},
  journal={arXiv preprint arXiv:2508.03983},
  year={2025}
}

@article{hubert,
  title={{HuBERT: Self-Supervised Speech Representation Learning by Masked Prediction of Hidden Units}},
  author={Hsu, Wei-Ning and Bolte, Benjamin and Tsai, Yao-Hung Hubert and Lakhotia, Kushal and Salakhutdinov, Ruslan and Mohamed, Abdelrahman},
  journal={IEEE/ACM Transactions on Audio, Speech, and Language Processing},
  volume={29},
  pages={3451--3460},
  year={2021},
  publisher={IEEE}
}

@inproceedings{audiomae,
  title={{Masked Autoencoders that Listen}},
  author={Huang, Po-Yao and Xu, Hu and Li, Juncheng and Baevski, Alexei and Auli, Michael and Galuba, Wojciech and Metze, Florian and Feichtenhofer, Christoph},
  booktitle={Proc. NeurIPS},
  year={2022},
  address={New Orleans},
}

@inproceedings{mert,
  title={{MERT: Acoustic Music Understanding Model with Large-Scale Self-Supervised Training}},
  author={Li, Yizhi and Yuan, Ruibin and Zhang, Ge and Ma, Yinghao and Chen, Xingran and Yin, Hanzhi and Xiao, Chenghao and Lin, Chenghua and Ragni, Anton and Benetos, Emmanouil and others},
  booktitle={Proc. ICLR},
  year={2024},
  address={Vienna},
}

@article{muq,
  title={{MuQ: Self-Supervised Music Representation Learning with Mel Residual Vector Quantization}},
  author={Zhu, Haina and Zhou, Yizhi and Chen, Hangting and Yu, Jianwei and Ma, Ziyang and Gu, Rongzhi and Luo, Yi and Tan, Wei and Chen, Xie},
  journal={IEEE Transactions on Audio, Speech and Language Processing},
  year={2025},
  volume={33},
  pages={3653-3664},
  publisher={IEEE}
}

@inproceedings{usad,
  title={{USAD: Universal Speech and Audio Representation via Distillation}},
  author={Chang, Heng-Jui and Bhati, Saurabhchand and Glass, James and Liu, Alexander H},
  booktitle={Proc. ASRU},
  year={2025},
  address={Hawaii}
}

@inproceedings{audioflamingo,
  title={{Audio Flamingo: A Novel Audio Language Model with Few-Shot Learning and Dialogue Abilities}},
  author={Kong, Zhifeng and Goel, Arushi and Badlani, Rohan and Ping, Wei and Valle, Rafael and Catanzaro, Bryan},
  booktitle={Proc. ICML},
  year={2024},
  address={Vienna}
}

@inproceedings{mmau-pro,
  title={{MMAU-Pro: A Challenging and Comprehensive Benchmark for Holistic Evaluation of Audio General Intelligence}},
  author={Kumar, Sonal and Sedl{\'a}{\v{c}}ek, {\v{S}}imon and Lokegaonkar, Vaibhavi and L{\'o}pez, Fernando and Yu, Wenyi and Anand, Nishit and Ryu, Hyeonggon and Chen, Lichang and Pli{\v{c}}ka, Maxim and Hlav{\'a}{\v{c}}ek, Miroslav and others},
  booktitle={Proc. AAAI},
  year={2026},
  address={Singapore}
}

@inproceedings{mmar,
  title={{MMAR: A Challenging Benchmark for Deep Reasoning in Speech, Audio, Music, and Their Mix}},
  author={Ma, Ziyang and Ma, Yinghao and Zhu, Yanqiao and Yang, Chen and Chao, Yi-Wen and Xu, Ruiyang and Chen, Wenxi and Chen, Yuanzhe and Chen, Zhuo and Cong, Jian and others},
  booktitle={Proc. NeurIPS},
  year={2025},
  address={San Diego}
}

@inproceedings{mmsu,
  title={{MMSU: A Massive Multi-Task Spoken Language Understanding and Reasoning Benchmark}},
  author={Wang, Dingdong and Li, Junan and Wu, Jincenzi and Yang, Dongchao and Chen, Xueyuan and Zhang, Tianhua and Meng, Helen},
  booktitle={Proc. ICLR},
  year={2026},
  address={Rio de Janeiro}
}

@inproceedings{sentencebert,
  title={{Sentence-BERT: Sentence Embeddings using Siamese BERT-Networks}},
  author={Reimers, Nils and Gurevych, Iryna},
  booktitle={Proc. EMNLP},
  year={2019},
  address={Hong Kong}
}

@inproceedings{desed,
  title={{Sound Event Detection in Domestic Environments with Weakly Labeled Data and Soundscape Synthesis}},
  author={Turpault, Nicolas and Serizel, Romain and Shah, Ankit Parag and Salamon, Justin},
  booktitle={Proc. DCASE Workshop},
  year={2019},
  address={New York}
}

@inproceedings{partialedit,
  title={{PartialEdit: Identifying Partial Deepfakes in the Era of Neural Speech Editing}},
  author={Zhang, You and Tian, Baotong and Zhang, Lin and Duan, Zhiyao},
  booktitle={Proc. Interspeech},
  year={2025},
  address={Rotterdam}
}

@article{audioflamingo-next,
  title={{Audio Flamingo Next: Next-Generation Open Audio-Language Models for Speech, Sound, and Music}},
  author={Ghosh, Sreyan and Goel, Arushi and Jayakumar, Kaousheik and Koroshinadze, Lasha and Anand, Nishit and Kong, Zhifeng and Gururani, Siddharth and Lee, Sang-gil and Kim, Jaehyeon and Aljafari, Aya and others},
  journal={arXiv preprint arXiv:2604.10905},
  year={2026}
}

@article{shi2026towards,
  title={{Towards Fine-grained Temporal Perception: Post-Training Large Audio-Language Models with Audio-Side Time Prompt}},
  author={Shi, Yanfeng and Cai, Pengfei and Liu, Jun and Gu, Qing and Jiang, Nan and Dai, Lirong and McLoughlin, Ian and Song, Yan},
  journal={arXiv preprint arXiv:2604.13715},
  year={2026}
}

\newpage

 




\vfill

\end{document}